\documentclass[aps,prl,twocolumn,eqsecnum,amssymb,amsmath,showpacs,a4paper, superscriptaddress]{revtex4-2}

\usepackage{graphicx}
\usepackage{amsfonts}
\usepackage{amsmath}
\usepackage{hyperref}
\usepackage{units}
\usepackage{color}
\usepackage{dcolumn}
\usepackage{bm}
\usepackage{float}
\usepackage{cleveref}
\usepackage[normalem]{ulem}
\usepackage{appendix}
\usepackage{mathtools}
\usepackage{esvect}

\usepackage[colorinlistoftodos, size=tiny, bordercolor=white]{todonotes}
\usepackage{comment}

\usepackage[subpreambles=true]{standalone}  
\usepackage{import} 

\graphicspath{ {images/} }
\usepackage{mathrsfs}
\usepackage{amssymb}
\usepackage{dsfont}
\usepackage{enumitem}
\usepackage{gensymb}
\usepackage{bm}

\crefname{section}{§}{§§}
\Crefname{section}{§}{§§}
\usepackage{mathtools}
\usepackage{chngcntr}
\counterwithout{equation}{section}

\usepackage{tikz} 
\usetikzlibrary{shapes,arrows,positioning,automata,backgrounds,calc,er,patterns}
\usepackage[compat=1.1.0]{tikz-feynman}

\makeatletter
\let\cat@comma@active\@empty
\makeatother

\newcommand{\cmmnt}[1]{\ignorespaces}

\newcommand{\be} {\begin{equation}}
\newcommand{\ee} {\end{equation}}
\newcommand{\bsub}{\begin{subequations}}
\newcommand{\esub}{\end{subequations}}
\newcommand{\bea}{\begin{eqnarray}}
\newcommand{\eea}{\end{eqnarray}}

\newcommand{\m}{\widetilde{m}}

\begin{document}

\title{Origin and evolution of the multiply-quantised vortex instability}

\author{Sam Patrick}
\affiliation{Department of Physics and Astronomy, University of British Columbia, Vancouver, British Columbia, V6T 1Z1, Canada}
\affiliation{Institute of Quantum Science and Engineering, Texas A\&M University, College Station, Texas, 77840, US}

\author{August Geelmuyden}
\affiliation{School of Mathematical Sciences, University of Nottingham, University Park, Nottingham, NG7 2RD, UK}

\author{Sebastian Erne}
\affiliation{School of Mathematical Sciences, University of Nottingham, University Park, Nottingham, NG7 2RD, UK}
\affiliation{Vienna Center for Quantum Science and Technology, Atominstitut, TU Wien, Stadionallee 2, 1020 Vienna, Austria}

\author{Carlo F. Barenghi}
\affiliation{Joint Quantum Centre Durham-Newcastle, School of Mathematics, Statistics and Physics, Newcastle University,
Newcastle upon Tyne, NE1 7RU, UK}

\author{Silke Weinfurtner}
\affiliation{School of Mathematical Sciences, University of Nottingham, University Park, Nottingham, NG7 2RD, UK}
\affiliation{Centre for the Mathematics and Theoretical Physics of Quantum Non-Equilibrium Systems, University of Nottingham, Nottingham, NG7 2RD, UK}

\date{\today}

\begin{abstract}
\noindent
We show that the dynamical instability of quantum vortices with more than a single quantum of angular momentum results from a superradiant 
bound state inside the vortex core. 
Our conclusion is supported by an analytic WKB calculation and numerical simulations of both  
linearised and fully non-linear equations of motion for a doubly-quantised vortex at the centre of a circular bucket trap. 
In the late stage of the instability, we reveal a striking novel behaviour of the system in the non-linear regime. 
Contrary to expectation, in the absence of dissipation the system never enters the regime of two well-separated phase defects described by Hamiltonian vortex dynamics. 
Instead, the separation between the two defects undergoes modulations which never exceed a few healing lengths, in which compressible kinetic energy and incompressible kinetic energy are exchanged. 
This suggests that, under the right conditions, pairs of vortices may be able to form meta-stable bound states. 
\end{abstract}

\maketitle
{\textbf{Introduction.}}---
A striking property of quantum fluids (superfluid helium, atomic Bose-Einstein condensates, polariton condensates, etc.) is that the circulation of the velocity $\mathbf{v}$ around a closed path $C$ is quantized \cite{feynman1955chapter} in units of $\kappa=2\pi\hbar/M$,
\begin{equation}
\oint_C \mathbf{v} \cdot d\mathbf{r}=\ell\kappa,
\end{equation}
where $M$ is the mass of the relevant boson, $\hbar$ is the reduced Planck's constant, and the integer $\ell$ is called the winding number.  
In most regions of fluid the circulation will be zero, but there may be points (in 2D) or lines (in 3D) where the wavefunction $\Psi$ vanishes, hence its phase is not defined and $\ell \neq 0$. 
Such topological defects (singularities),  normally surrounded by circular (in 2D) or tubular (in 3D) regions of depleted density, are called quantum vortices. 
The topological nature of these vortices deeply affects the possible flow patterns (vortex lattices, turbulence, etc.)

Experiments \cite{shin2004dynamical,isoshima2007spontaneous,okano2007splitting} show that a multiply-quantised vortex (MQV), i.e. a vortex with $\ell>1$, will spontaneously decay into a cluster of singly-quantised vortices (SQVs), each with $\ell=1$.  
This tendency is usually justified on the grounds 
that, for a given angular momentum, a cluster of SQVs is energetically favourable as compared with a MQV \cite{barenghi2016primer}.
Hence, in a dissipative scenario, where the system relaxes into the lowest energy state, an MQV will naturally evolve into a cluster of SQVs.
In non-dissipative systems, however, the decay can still occur due to a dynamical instability \cite{Pu1999} arising from the coupling of the MQV to surrounding phonons.

The instability of MQVs acquires additional significance if we note that, under certain conditions, there are analogies between vortices and rotating black holes \cite{Torres2019}.
It has recently been argued \cite{giacomelli2020ergoregion} that the dynamical instability is related to the existence of an ergoregion, a notion from black hole physics which implies superradiant amplification of waves in a particular frequency range \cite{brito2020superradiance}.
Superradiance arises not only around rotating black holes but in a wide range of systems, e.g.
draining vortices \cite{torres2017rotational} and optical vortex 
beams \cite{braidotti2021measurement}.
However, unbounded growth can occur if there is a mechanism for trapping superradiant modes in the system, eventually driving it into the non-linear regime, like e.g. black hole bomb instabilities \cite{dolan2007instability}.

In this Letter we study the evolution of an $\ell=2$ MQV in a bucket trap. 
Our analytic WKB prediction confirms the superradiant character of the dynamical instability in the initial linear regime \cite{giacomelli2020ergoregion}. 
By solving the full non-linear equations, we also reveal a remarkable recurrent behaviour of the instability at later times: a modulation in which incompressible kinetic energy and compressible kinetic energy are periodically exchanged, and the two phase defects, which we call proto-vortices, move in and out while rotating in close proximity to each other, unable either to form two fully-fledged separated SQVs or to merge back into an MQV. 
This novel time-dependent state is the limiting configuration of two parallel quantum vortices at close distances comparable to the healing length.
We also show that the new time-dependent state can be captured by a simple two-mode model representing the proto-vortex separation and the dynamically unstable phonon mode.
The role of dissipation then becomes clear: dissipation prevents the coherent reabsorption of phonons, allowing the proto-vortices to spiral out and separate, becoming well-defined quantum vortices each surrounded by their own core regions.

\begin{figure*} 
\centering
\includegraphics[width=\linewidth]{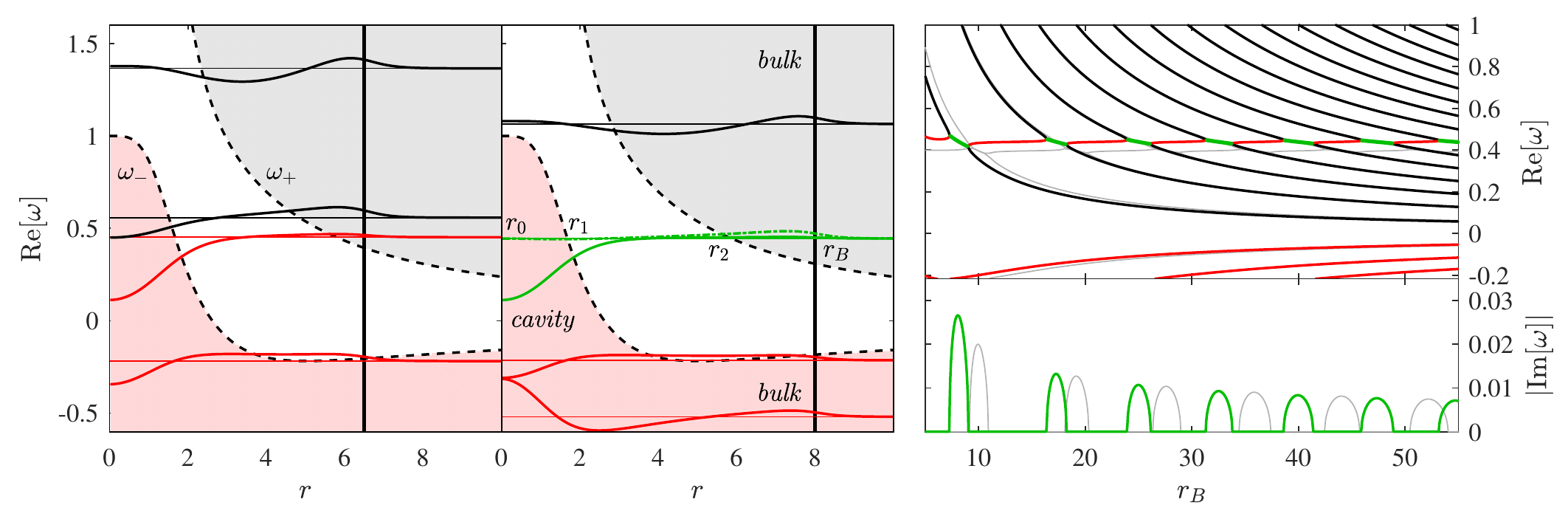}
\caption{\textbf{Left panel}: Oscillation frequencies and the associated eigenmodes as a function of radius. Horizontal lines are the real part of the eigenvalues $\mathrm{Re}[\omega]$ and the superimposed thick lines are the relative density eigenmodes $\delta\rho/\rho=u_++u_-$, which are solutions of the BdG \eqref{BdG} for $m=\ell=2$. The solid vertical line represents the trap size $r_B=6.5$. Dashed black curves are the WKB potentials $\omega_\pm$ defined in \eqref{potentials}, which separate the grey (positive norm), white (evanescent) and pink (negative norm) regions. Modes with $\mathcal{N}>0$ ($\mathcal{N}<0$) are coloured black (red). \textbf{Central panel}: the same for $r_B=8$. The unstable mode, whose real (imaginary) part is shown as solid (broken) green line, results from the coupling of two nearby modes in the left panel. The complex conjugate of this mode (not shown) is a decaying solution. Also illustrated are the turning points $r_i$ defined below \eqref{potentials}. \textbf{Right panel}: the eigenvalues $\omega$ as a function of $r_B$, with the real (imaginary) part in the upper (lower) panel. BdG solutions are shown as thick lines and follow the same colour scheme as the previous planels. Solutions to the WKB condition \eqref{cot} are shown as grey in the background and, for high frequencies, are indistinguishable from the BdG results.}
\label{fig:1}
\end{figure*}

{\textbf{Vortex states.}}---
We consider the dimensionless two-dimensional Gross-Pitaevskii equation (GPE),
\begin{equation} \label{GPE}
i\partial_t\Psi = \left(-\tfrac{1}{2}\nabla^2 + V(\mathbf{x})-1 +|\Psi|^2\right)\Psi,
\end{equation}
where lengths are measured in units of $\xi \equiv \hbar/\sqrt{M\mu}$
(the healing length), time in units of $\tau \equiv \hbar/\mu$,
and density $\vert \Psi\vert^2$ in units of $\mu/g$. 
Here, $\mu$ is the chemical potential, $M$ is the atomic mass and $g$ is the 2D interaction strength.
We work in polar coordinates $\mathbf{x}=(r,\theta)$.
The condensate is confined by a circular bucket-potential of the form,
\begin{equation} \label{trap}
V(r) = \frac{V_0}{1+(V_0-1)e^{a(r_B-r)}},
\end{equation}
where $r_B$ is the trap size and $V_0$ and $a$ determine the steepness of the bucket wall at $r_B$. We choose $a=V_0=5$, although our results are essentially independent of this choice provided the wall at $r_B$ is steep.

Using the Madelung representation of the condensate wavefunction $\Psi=\sqrt{\rho}e^{i\Phi}$, the stationary GPE has vortex solutions with velocity $\mathbf{v}\equiv\bm{\nabla}\Phi=\ell/r\, \vv{\mathbf{e}}_\theta$, where $\bm{\nabla}$ is the 2D gradient operator. Here we focus on the doubly winded vortex with $\ell=2$. The corresponding density profile can then be obtained by substituting $\Phi=\ell\theta$ in \eqref{GPE} and solving numerically for $\rho$.

{\textbf{BdG equation.}}---
Since $V(\mathbf{x})$ is independent of $t$ and $\theta$, linear fluctuations $\delta\psi$ of the condensate wavefunction can be decomposed into separate frequency $\omega$ and azimuthal $m$ components,
\begin{equation}
\begin{pmatrix}
\delta\psi \\ \delta\psi^*
\end{pmatrix} = \int^\infty_{-\infty}\frac{d\omega}{2\pi}\sum_{m=-\infty}^\infty e^{im\theta-i\omega t} \begin{pmatrix}
u_+e^{+i\ell\theta-it} \\ u_-e^{-i\ell\theta+it}
\end{pmatrix},
\end{equation}
where we write $u_\pm=u_\pm(\omega,m,r)$ for brevity.
The fluctuations can then be described the state $|U\rangle = (u_+,u_-)^\mathrm{T}$ which obeys the Bogoliubov-de Gennes (BdG) equation,
\begin{equation} \label{BdG}
\begin{split}
    \widehat{L}|U\rangle = & \ \omega |U\rangle, \qquad \widehat{L} = \begin{pmatrix}
D_+ & \rho \\ -\rho & -D_-
\end{pmatrix}, \\
D_\pm = -\frac{1}{2}\bigg[\partial_r^2+ & \ \frac{1}{r}\partial_r -\frac{(m\pm\ell)^2}{r^2}\bigg] + V(r) + 2\rho -1.
\end{split}
\end{equation}
The BdG conserves a quantity called the norm,
\begin{equation}
    \mathcal{N} = \int d^2\mathbf{x}\left(|u_+|^2-|u_-|^2\right),
    \end{equation}
which is related to the mode energy by a factor of $\omega$.
Solutions to \eqref{BdG} are obtained by diagonalising $\widehat{L}$ to obtain the eigenvalues $\omega$ and eigenfunctions $|U\rangle$.
In the right panel of Fig.~\ref{fig:1}, we display $\omega$ as a function of the trap size $r_B$ and indicate with colour the sign of $\mathcal{N}$.
The instability (a zero norm solution) results from the coupling of a positive norm mode to a negative norm one, which (for $\ell=2$) can only occur for $m=2$.
For particular system sizes, this coupling is suppressed and the instability is absent.
In the limit that $r_B\to\infty$, the density of $\mathcal{N}>0$ states becomes a continuum and the coupling always occurs \cite{giacomelli2020ergoregion}.
Note that the unstable growing mode ($\mathrm{Im}[\omega]>0$) is always accompanied by its complex conjugate, which corresponds to a stable decaying mode ($\mathrm{Im}[\omega]<0$).
Examples of the relative density eigenfunctions $\delta\rho/\rho=u_++u_-$ are also shown on Fig.~\ref{fig:1}.


{\textbf{WKB method.}}---
Deeper insight into these results can be obtained via a WKB approximation, wherein the fluctuations are assumed to behave locally like plane waves. This provides local scattering information about the waves.
Inserting the Ansatz $u_\pm\sim A_\pm(r) \exp(i\int p(r)\,dr)$ into \eqref{BdG} and neglecting derivatives of the amplitudes, we obtain the dispersion relation,
\begin{equation} \label{disprel}
\Omega^2 = \rho k^2 + k^4/4, \qquad k = (p^2+\m^2/r^2)^\frac{1}{2},
\end{equation}
where $\Omega=\omega-m\ell/r^2$ is the frequency in a frame co-moving with the vortex and the effective azimuthal number $\m$ in the expression for $k$ is given by,
\begin{equation}
    \m^2=m^2+\ell^2+2r^2(\rho+V(r)-1).
\end{equation}
Since Eq.~\eqref{disprel} is quadratic in $p^2$, the dispersion relation will have two pairs of solutions.
One of the pairs corresponds to solutions which are evanescent throughout the system and must be discarded for the solution to be regular at $r=0$.
The second pair corresponds to radially in- and out-going plane waves (as determined by the sign of the radial group velocity $\vv{\mathbf{e}}_r\cdot\bm{v}_g\equiv\partial_p\omega$) far from the vortex core and are the relevant ones for the discussion of the instability.

When these modes are propagating, scattering occurs at turning points $r_i$ (locations where $\vv{\mathbf{e}_r}\cdot\bm{v}_g(r_i)=0$ which occurs for $p=0$). Substituting $p=0$ back into \eqref{disprel} yields two curves,
\begin{equation} \label{potentials}
\omega_\pm = \frac{m\ell}{r^2} \pm \sqrt{\rho\frac{\m^2}{r^2}+\frac{\m^4}{4r^4}},
\end{equation}
which determine the locations of the turning points $r_i$ through the relations $\omega=\omega_\pm(r_i)$, see Fig.~\ref{fig:1} for an example.
Between the two curves, the modes are evanescent, whereas above $\omega_+$ (below $\omega_-$) they are propagating and have $\Omega>0$ ($\Omega<0$).
In the WKB approximation, the sign of $\Omega$ coincides with the sign of $\mathcal{N}$ in that region.
Hence, a frequency which intersects both $\omega_+$ and $\omega_-$ will tunnel from the positive norm branch of the dispersion to the negative norm one.
Due to norm conservation, such a mode will be amplified (superradiantly) each time it scatters with a turning point.

The possibility of having instabilities arises when amplified modes are reflected back into the system where they are further amplified.
Such a scenario occurs for vortices since there is a region inside the vortex core (a cavity, see Fig.~\ref{fig:1} central panel) where these modes can become trapped.
Broadly speaking, there will be an instability if there is a bound state inside the vortex core (cavity mode) which couples to one of the normal modes outside the vortex in the bulk (phonons).
Such a coupling is only possible in the range of frequencies which probe the cavity (the cavity band).
In an infinite system, the spectrum of phonons is continuous, hence such a coupling is guaranteed.
This is not the case for a finite sized system since the phonon spectrum becomes discrete.

To evaluate the normal mode frequencies with the WKB method, we perform a scattering computation analogous to finding normal mode frequencies of the Schr\"odinger equation for a potential containing two wells.
The full procedure is detailed in our companion paper \footnote{\label{companion}In preparation.} and leads to the following condition for frequencies in the cavity band,
\begin{equation} \label{cot}
4\cot(S_{01})\cot(S_{2B}+\pi/4) = \exp(-2S_{12}),
\end{equation}
where $S_{ij}(\omega) = \int^{r_j}_{r_i}|p(\omega)|dr$ is the phase integral between the turning points, see Fig.~\ref{fig:1}.
In deriving this condition, the frequency is assumed real.
For a small imaginary part, the extension to complex frequencies proceeds via $S(\omega)\simeq S(\mathrm{Re}[\omega])+i\mathrm{Im}[\omega]\partial_\omega S|_{\mathrm{Re}[\omega]}$ \footnote{Note that since the WKB method fails close to $r_B$ where $V$ in \eqref{trap} varies rapidly, we approximate the integral $S_{2B}$ by assuming that $\rho$ has the same form as that in the infinite system. 
Imposing a Neumann boundary condition at $r_B$ gives the $\pi/4$ term in \eqref{cot}.}.

When the zeros of the two cotangent functions are well separated, \eqref{cot} describes two sets of solutions; cavity modes satisfy $\cos S_{01}\simeq 0$, whereas phonons obey $\cos(S_{2B}+\pi/4)\simeq 0$.
However, if members of these two sets are close enough in frequency, a coupling occurs resulting in two zero norm modes which are a complex conjugate pair.
Solutions to \eqref{cot} are shown in light grey on the right panels of Fig.~\ref{fig:1}.
The high frequency solutions (relative to the trap size) agree exceptionally well with the BdG results.
The discrepancy of the cavity mode frequency results from the fact that this mode occupies a region where the density varies rapidly, thereby increasing the error of WKB. This is also the reason why WKB fails to match quantitatively the locations of the stability windows.
Nonetheless, the approximation captures all the features of the spectrum, which is enough to validate the interpretation gained with this method.

The connection to superradiance can be seen explicitly by looking in the large system limit $r_B\to\infty$ \footnote{See our companion paper [13] for a comment on the difference between $r_B\to\infty$ and a system which is truly infinite, i.e. has open boundary conditions}.
In that case, the condition in \eqref{cot} reduces to,
\begin{equation}
\cos S_{01}(\mathrm{Re[}\omega]) = 0, \qquad \mathrm{Im[}\omega] = -\frac{\log|\mathcal{R}|}{2\partial_\omega S_{01}}\bigg|_{\mathrm{Re[}\omega]},
\end{equation}
where $|\mathcal{R}|=(1+e^{-2S_{12}}/4)/(1-e^{-2S_{12}}/4)$ is the local reflection coefficient associated with the tunnelling between the bulk and the cavity.
The factor on the denominator is negative since the cavity modes have $\mathcal{N}<0$ in that region.
Hence, the reason these modes are unstable is a direct consequence of superradiance, i.e. $|\mathcal{R}|>1$.

\begin{figure*} 
\centering
\includegraphics[width=\linewidth]{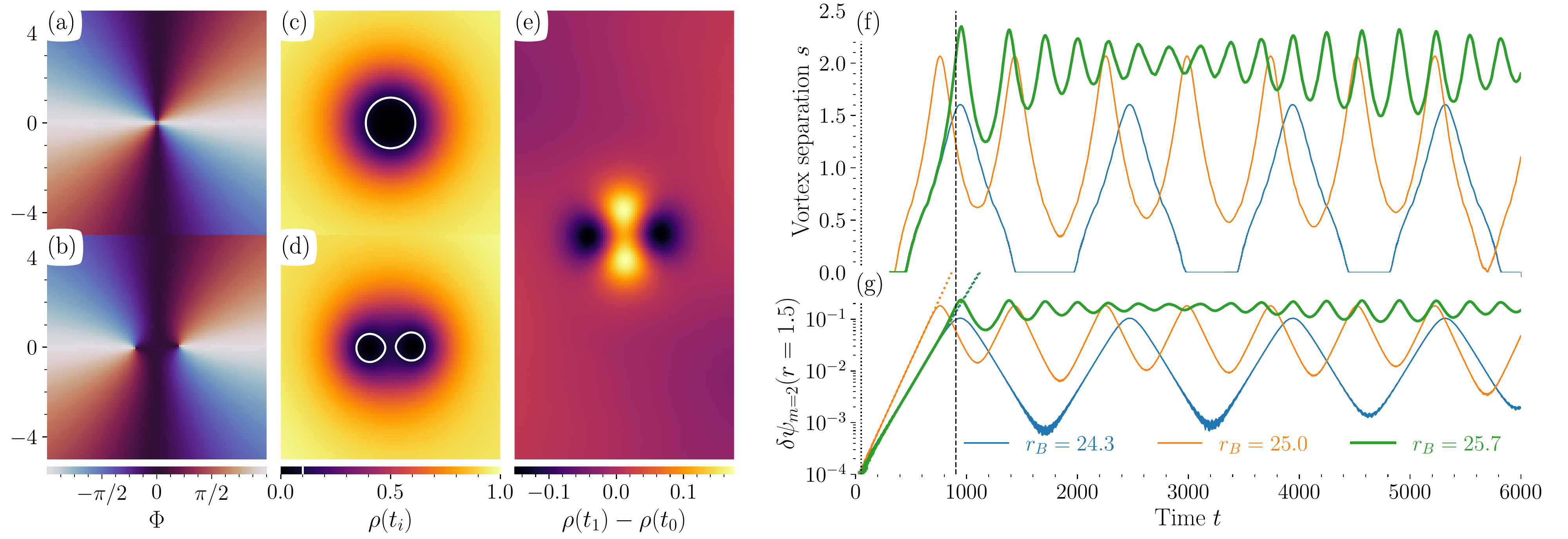}
\caption{The splitting of the initial MQV into two separate singularities (proto-vortices).
Panels (a) and (b) show the phase $\Phi$ for $r_B = 25$ at $t = 50$ and $t = 904$ respectively.
Panels (c) and (d) show the density $\rho$ at the same times; here white lines are surfaces of $\rho = 1/10$. 
Panel (e) displays the difference between the density at $t=50$ (c) and the density at $t=904$ (d): the dominant $m = 2$ mode is apparent, whose troughs coincide with the locations of the two proto-vortex centres $\mathbf{x}_1$ and $\mathbf{x}_2$. The separation $s=|\mathbf{x}_1-\mathbf{x}_2|$ between the two singularities is shown to oscillate with $t$ in panel (f) for three different values of the trap size $r_B$.
Finally, panel (g) shows the time evolution amplitude of the $m=2$ mode inside the cavity. 
The initial growth rates agree with the prediction of our linear analysis, shown as dotted lines. Note that the blue and green dotted lines are essentially overlapping since the growth rates are the same.
The non-linear evolution after the initial stage ($t\gtrsim1000$) consists of a cycle between growth and decay. }
\label{fig:2}
\end{figure*}

{\textbf{Non-linear evolution.}}---
To verify the existence of the instability in the full non-linear system, we simulate the full GPE in \eqref{GPE}.
Simulations are prepared with the following wave function,
\begin{equation}
    \Psi(\mathbf{x},t=0)=\sqrt{\rho_{\ell=2}(r)}e^{2i\theta}(1+\,\delta\Psi).
\end{equation}
The first term corresponds to the doubly-quantised vortex background and $\delta\Psi = \varepsilon(u_+e^{2i\theta}+u_-^*e^{-2i\theta})$ is the seed for the instability where $u_\pm$ are the unstable solutions of \eqref{BdG} for $m=2$.
The constant $\varepsilon$ is an overall amplitude which keeps the linear perturbation in the initial condition small; we choose $\varepsilon=10^{-3}$, although provided $\varepsilon$ is small enough, the only change to our results is the length of time before the initial instability becomes $\mathcal{O}(1)$. This state is taken as the initial condition for the evolution dictated by \eqref{GPE}, for which we use a two-split Fourier-Spectral scheme as outlined in \cite{Javanainen_2006}, see also the Supplemental Material (SM) \cite{supp} for details. In Fig.~\ref{fig:2}, we display snapshots of $\Psi$ for $r_B=25$ and evolution of the vortex separation for three different values of $r_B$.

\begin{figure} 
\centering
\includegraphics[width=\linewidth]{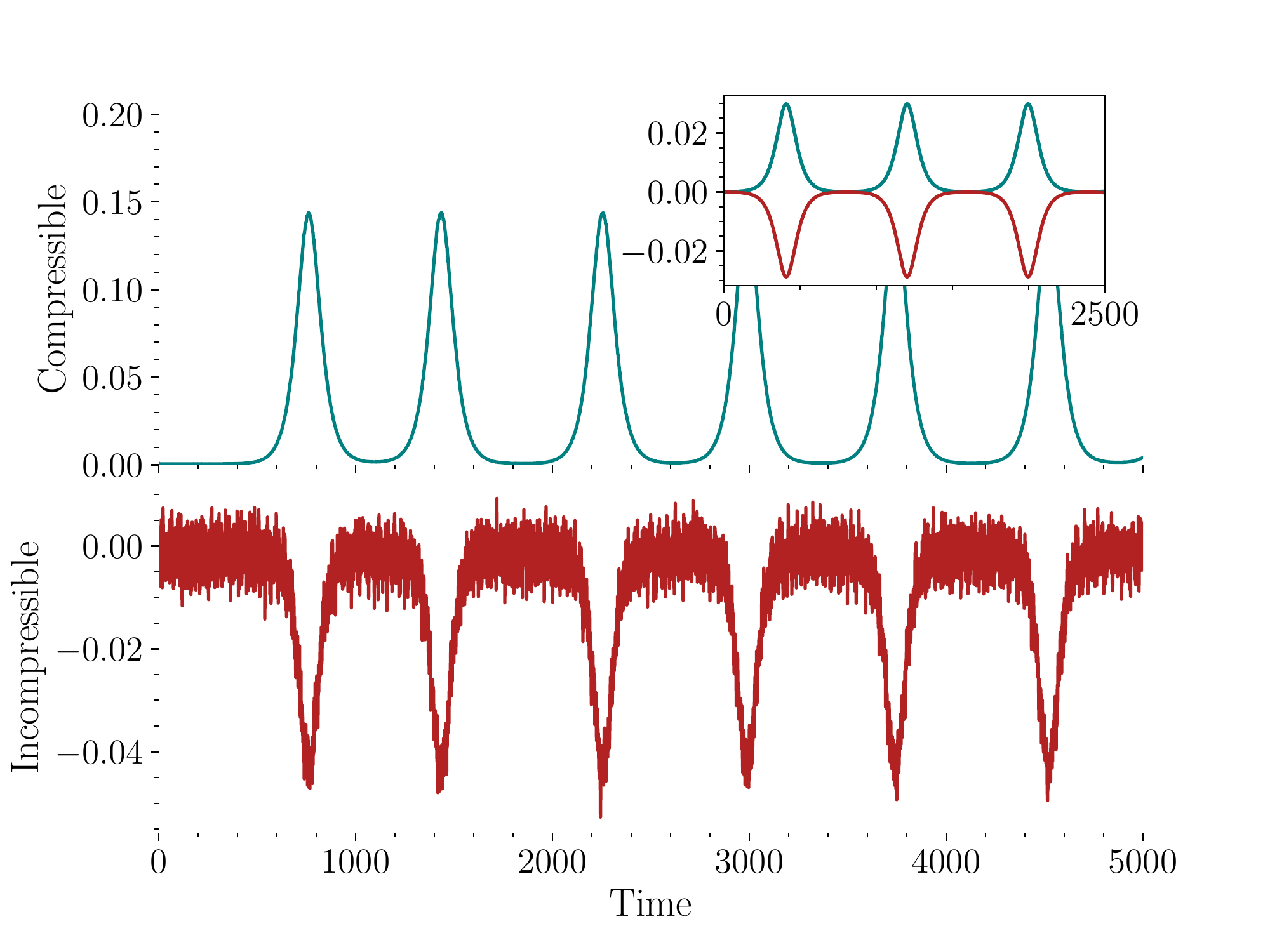}
\caption{
Time evolution of the compressible (green line) and incompressible (red line) kinetic energies associated with phonons and proto-vortices respectively. Note that the increase of the compressible kinetic energy corresponds to the decrease of the incompressible kinetic energy.
This supports our interpretation of the observed modulations being driven by a back-and-forth energy exchange between phonons and proto-vortices. The inset shows the energies of our simple oscillator model described in the SM \cite{supp}, i.e. $H_+$ (green line) and $H_-$ (red line), which captures the behaviour of proto-vortices and phonons. For the model, we used the parameters $\Omega=1$, $\sigma=1/50$, $g=1/2    00$, $c_+=0$, and $c_-=2$.
}
\label{fig:energy}
\end{figure}

We find that the instability predicted by the linear analysis is present at early times, see panel (g) in Fig.~\ref{fig:2} for $t\lesssim1000$ \footnote{We have also confirmed that unstable modes are absent in the stable windows predicted by Fig.~\ref{fig:1}}.
Whilst the unstable mode is growing, we observe that the initial $\ell=2$ phase singularity splits into two $\ell=1$ singularities (panels (b) and (d) of Fig.~\ref{fig:2}), each guided by a trough of the unstable $m=2$ mode. Indeed, it is precisely the growth of the negative energy part of the mode (occupying the core of the original vortex) which causes the two resulting singularities to spiral outward, since this lowers the energy of the overall vortex configuration \cite{barenghi2016primer}.
Surprisingly, we find that once the separation between the singularities reaches approximately $2$ or $3$ healing lengths, they begin to spiral back inwards.
Since the two singularities occupy the same region of depleted density, we refer to them as proto-vortices, as the name vortex is usually understood as a singularity embedded in its own low density region (the vortex core).
During the inward stage, the $m=2$ waveform matches that of the linear decaying mode discussed in the caption of Fig.~\ref{fig:2}.
This mode can be thought of as describing the re-absorption of a phonon by the proto-vortex pair.
Eventually, the decay halts and growth resumes, causing a modulation of the separation.
This cycle continues for the duration of our simulations, trading kinetic energy back and forth. The kinetic energy of the system, in fact, has two contributions \cite{Nore1997}: compressible 
(arising from phonons) and incompressible (arising from the proto-vortices), see SM \cite{supp} for the definitions.
Fig.~\ref{fig:energy} clearly shows that when the former increases, the latter decreases.

We remark that during the late-stage of the instability, the two proto-vortices remain confined inside the region of suppressed density (see Fig.~\ref{fig:2} panel (d)). Moreover, the maximum value of the density perturbation $|\delta \Psi|$ during the evolution is only about 0.2. This
suggests that the full non-linear dynamics can indeed be described perturbatively
about the original $\ell = 2$ vortex background, contrasting the expectation that a MQV should decay non-perturbatively into a cluster of well-separated SQVs.

In the SM \cite{supp} we describe a simple dynamical system consisting 
of two oscillators $\mathbf{X}_\pm$ with opposite sign energies $H_\pm=\pm(\frac{1}{2}\dot{\mathbf{X}}^2_\pm+V_\pm)$, which describes the essential interaction between proto-vortices and phonons.
The potential and interaction energies are $V_\pm = \frac{1}{2}(\Omega^2\pm\sigma)\mathbf{X}_\pm^2-\frac{1}{4}c_\pm\mathbf{X}^4_\pm$ and $V_\mathrm{int}=g\mathbf{X}_+\cdot\mathbf{X}_-$.
We show how this system can be solved in the limit of large $\Omega$ to reveal the signature switch between exponential growth and decay seen in our simulations, as shown in the inset of Fig.~\ref{fig:energy}.
The characteristic feature which allows this is that the negative energy cavity mode (which results in the formation of proto-vortices) couples only to a single phonon in the bulk, and couplings to all other phonons are considered negligible.

The essential physics is captured in the following qualitative argument.
Whilst the instability is growing, the non-linearity in the GPE reduces the frequency of the cavity mode until it can no longer couple to the phonon that rendered it unstable. After decoupling, the phonon will oscillate faster than the cavity mode and the phase difference between the two will change.
Once $\int (\omega_\mathrm{ph}-\omega_\mathrm{vor})dt\simeq\pi$, the waveforms of the two modes will be just right to produce the decaying mode once they recombine.
For this argument to hold, the density of phonon states in Fig.~\ref{fig:1} must be sufficiently low that the cavity mode cannot couple to any phonons of lower frequency as it evolves.
Hence we would not expect modulations of the proto-vortex separation to occur in a very large system where the density of phonon states is effectively continuous.
This being said, we have found that modulations can persist for trap sizes up to at least $r_B=47$.

One might assume that this behaviour is highly sensitive to initial conditions and that any small perturbation might be enough to destroy the effect. Whilst this seems to be the case when the vortex is placed far from the centre of the trap \cite{okano2007splitting}, we have found that our modulations persist when even if the vortex is displaced from the origin by a few healing lengths.

Finally, we have checked what happens when damping is added to the dynamics. We have found that energy is slowly removed from the system and, at late times, we recover the usual Hamiltonian point vortex behaviour. Before this, however, the cavity mode (whose frequency decreases monotonically) will couple to any available lower frequency phonons, causing sudden modulations of the vortex separation as the singularities drift apart (see the SM \cite{supp} for an example).

\textbf{Conclusion.}---
We have studied the instability of a doubly-quantised $\ell=2$ vortex using three distinct methods: a linear BdG stability analysis, a WKB approximation and a fully nonlinear numerical simulation of the GPE.
The WKB method allowed us to identify the cause of the instability as a superradiant bound state inside the vortex core.
We then confirmed that the instability predicted in the linear equations was also present in the full GPE dynamics.
Quite unexpectedly, we found that, whilst the instability is present at early times (and will cause the singularities to separate) the non-linearity in the GPE pushes the proto-vortices back together once they reach a critical separation, resulting in a modulation of their separation.
Whilst it was already predicted that instabilities can be suppressed in certain trap geometries, e.g. \cite{okano2007splitting,giacomelli2020ergoregion}, it was not known that an unstable vortex state could do something other than decay into a well separated pair of SQVs.
The observed modulations of the separation between singularities are suggestive that, under the right conditions, co-rotating vortex pairs may be able to form meta-stable bound states.
One possible interpretation of this is that each of the proto-vortices can be trapped inside the core of the other.
A consequence of this is that our system never enters the regime where one can apply Hamiltonian vortex dynamics \cite{newton2002n}, since this requires that the vortex separation be much larger than the healing length.
It would be interesting to see whether this behaviour extends to more general scenarios e.g. clusters of vortices.

\textbf{Acknowledgements.}---
SP acknowledges support from Natural Science and Engineering Research Council (Grant 5-80441 to W. Unruh) and would also like to thank the IQSE, Texas A\&M University, and grants ONR (Award No. N00014-20-1-2184) and NSF (Grant No. PHY-2013771) for support and an intellectually stimulating environment while part of this work was done.
CB and SW acknowledge support provided by the Science and Technology Facilities Council on Quantum Simulators for Fundamental Physics (ST/T00584X/1 and ST/T006900/1) as part of the Quantum Technologies for Fundamental Physics programme.
SW acknowledges support provided by the Leverhulme Research Leadership
Award (RL-2019 - 020), the Royal Society University
Research Fellowship (UF120112) and the Royal Society
Enhancement Grant (RGF/EA/180286), and partial support by the Science and Technology Facilities Council (Theory Consolidated Grant ST/P000703/1).
AG and SW acknowledge support provided by the Royal Society Enhancement Grant (RGF/EA/181015).
SE and SW acknowledge support from the EPSRC Project Grant
(EP/P00637X/1).
SE acknowledges partial support
through the Wiener Wissenschafts- und TechnologieFonds (WWTF) project No MA16 - 066 (``SEQUEX''),
and funding from the European Union's Horizon 2020
research and innovation programme under the Marie
Sklodowska-Curie grant agreement No 801110 and the
Austrian Federal Ministry of Education, Science and Research (BMBWF) from an ESQ fellowship.

\bibliography{new_biblio.bib}
\bibliographystyle{apsrev4-2}

\newpage

\appendix
\begin{center}
    \large{\textbf{Supplemental Material}}
\end{center}

\section{1.~Simulations}
\noindent The numerical simulations of vortex decay presented in Fig.~\ref{fig:2} is performed in three steps: (1) Preparation of the initial state, (2) Time evolution using the GPE \eqref{GPE} and (3) mode extraction. 

For a given trap potential $V(r)$ of the form \eqref{trap}, the initial state is prepared by first estimating the lowest energy configuration for a central $\ell=2$ vortex. This is done by fixing the phase $\Psi\equiv \sqrt{\rho}e^{i\ell\theta}$ and evolving the modulus $\sqrt{\rho}$ in imaginary time $\tau \equiv it$, i.e.
\begin{equation}
    \partial_\tau \sqrt{\rho} 
    = \left(\frac{1}{2}\partial_r^2
    +\frac{1}{2r} \partial_r
    - \frac{\ell^2}{2r^2} - V(r) + 1 - \rho\right)\sqrt{\rho}.
\end{equation}
The resulting density $\rho$ is inserted into a finite-difference matrix formulation of the BdG equation \eqref{BdG}. Numerically solving for the eigenmodes and selecting the solution $|U\rangle = (u_+,u_-)^T$ with the largest imaginary part, allows for the construction of the initial state,
\begin{equation}
    \Psi_0 = \sqrt{\rho_0(r)}e^{i\ell\theta}\left[1+\varepsilon u_+(r)e^{im\theta} + \varepsilon u_-^*(r)e^{-im\theta}\right].
    \label{supp:eq:initialState}
\end{equation}
Here $\varepsilon \ll 1$ is the initial amplitude of the unstable mode. For the data presented in Fig.~\ref{fig:2}, the values used are $\varepsilon = 10^{-3}$, $m=2$ and $\ell=2$.

The state $\Psi_0$ serves as the initial state for the full GPE simulation. Here, the cartesian plane is discretized into $N\times N$ linearly spaced mesh of pixels with separation of $\Delta l$ in each dimension. The time evolution proceeds in discrete timesteps of duration $\Delta t$ using a Fourier split operator method (see \cite{Javanainen_2006} for further details), which amounts to,
\begin{equation}
    \Psi(\mathbf{r},t+\Delta t)
    \simeq e^{\frac{i\Delta t}{2}\nabla^2} e^{-i\Delta t(V-1+|\Psi|^2)} \Psi(\mathbf{r},t).
    \label{supp:eq:timestep}
\end{equation}
Provided that the boundary $x,y = \pm \frac{1}{2}(N-1)\Delta l$ of the simulation domain is well outside the potential boundary $r_B$, the wavefunction $\Psi$ is sufficiently periodic for a Fourier-spectral evaluation of the exponentiated Laplacian in \eqref{supp:eq:timestep}. In the simulation presented in Fig.~\ref{fig:2}, the values $N=768$, $\Delta l=1/10$ and $\Delta t = 10^{-3}$ were used along with a potential of the form \eqref{trap} with $a=5$, $V_0=5$, $r_B = 25$.
Armed with the initial state and the discretization scheme outlined above, we perform $8192000=16384\times 500$ timesteps of which every $500^\mathrm{th}$ frame is stored for later processing. The result is a collection $\Psi(x_i,y_j,t_k)\in \mathbb{C}^{N^2\times N_t}$. 

To extract the evolution of the unstable mode, the wavefunction $\Psi(x_i,y_j,t_k)$ is transformed to polar coordinates $\Psi(r_i,\theta_j,t_k)$ and Fourier transformed in the azimuthal direction to give $\Psi(r_i,m_j,t_k)$ where $m_j$ is the $j^\mathrm{th}$ azimuthal component. The imaginary part $\text{Im}[\omega]$ of the frequency of the unstable mode is found by performing a log-linear fit to $|\Psi|(r_i,m_j,t_k)$ at $r_i=1.5$ and $m_j=2$. The real part $\text{Re}[\omega]$ is computed (at the same points) using a temporal Fourier transform.
 
The energy,
\begin{equation}
     E = 
     \int d^2 \mathbf{x}\left(
     \underbrace{\frac{1}{2}|\nabla \sqrt{\rho}|^2}_{E_\mathrm{qnt}} 
     + 
     \underbrace{\frac{1}{2} |\sqrt{\rho}\nabla \Phi|^2}_{E_\mathrm{kin}}
     + 
     \underbrace{V\rho}_{E_\mathrm{pot}}
     + 
     \underbrace{\frac{1}{2}\rho^2}_{E_\mathrm{int}}
     \right)
\label{eq:EnergyDecomposition}
\end{equation}
 associated with a state $\Psi = \sqrt{\rho}e^{i\Phi}$ in the GPE may be decomposed into a quantum energy $E_\mathrm{qnt}$, kinetic energy $E_\mathrm{kin}$, trap energy $E_\mathrm{pot}$, and interaction energy $E_\mathrm{int}$. As first proposed by Nore et al. \cite{nore1997kolmogorov, nore1997decaying}, the kinetic energy $E_\mathrm{kin}$ may be further split into a compressible part $E_\mathrm{kin}^{c}$ and an incompressible part $E_\mathrm{kin}^{i}$. Defining $\mathbf{u}\equiv \sqrt{\rho}\nabla \Phi$ and introducing $\mathbf{u}\equiv\mathbf{u}_c+\mathbf{u}_i$ with $\nabla \cdot \mathbf{u}_i = 0$, the two components of the kinetic energy takes the form 
 $E_\mathrm{kin}^{i} = \int d^2\mathbf{x} \frac{1}{2} |\mathbf{u}_i|^2$ and $E_\mathrm{kin}^{c} = \int d^2\mathbf{x} \frac{1}{2} |\mathbf{u}_c|^2$. Numerically, such a decomposition may be obtained from the realisation that if $\mathcal{F}$ denotes a spatial Fourier transform and $\mathbf{k}$ the corresponding wave vector, then $\mathbf{u}_c$ is nothing but the projection of $\mathbf{u}$ onto $\mathbf{k}$, i.e.
 \begin{equation}
     \mathbf{u}_c = \mathcal{F}^{-1}\left[\frac{\mathbf{k}(\mathbf{k}\cdot \mathcal{F}\mathbf{u})}{|\mathbf{k}|^2} \right],
 \end{equation}
 where, in the absence of a mean flow, the $\mathbf{k}=0$ component may be ignored to avoid zero-division. 
 
\section{2.~Damping}
\begin{figure} 
\centering
\includegraphics[width=\linewidth]{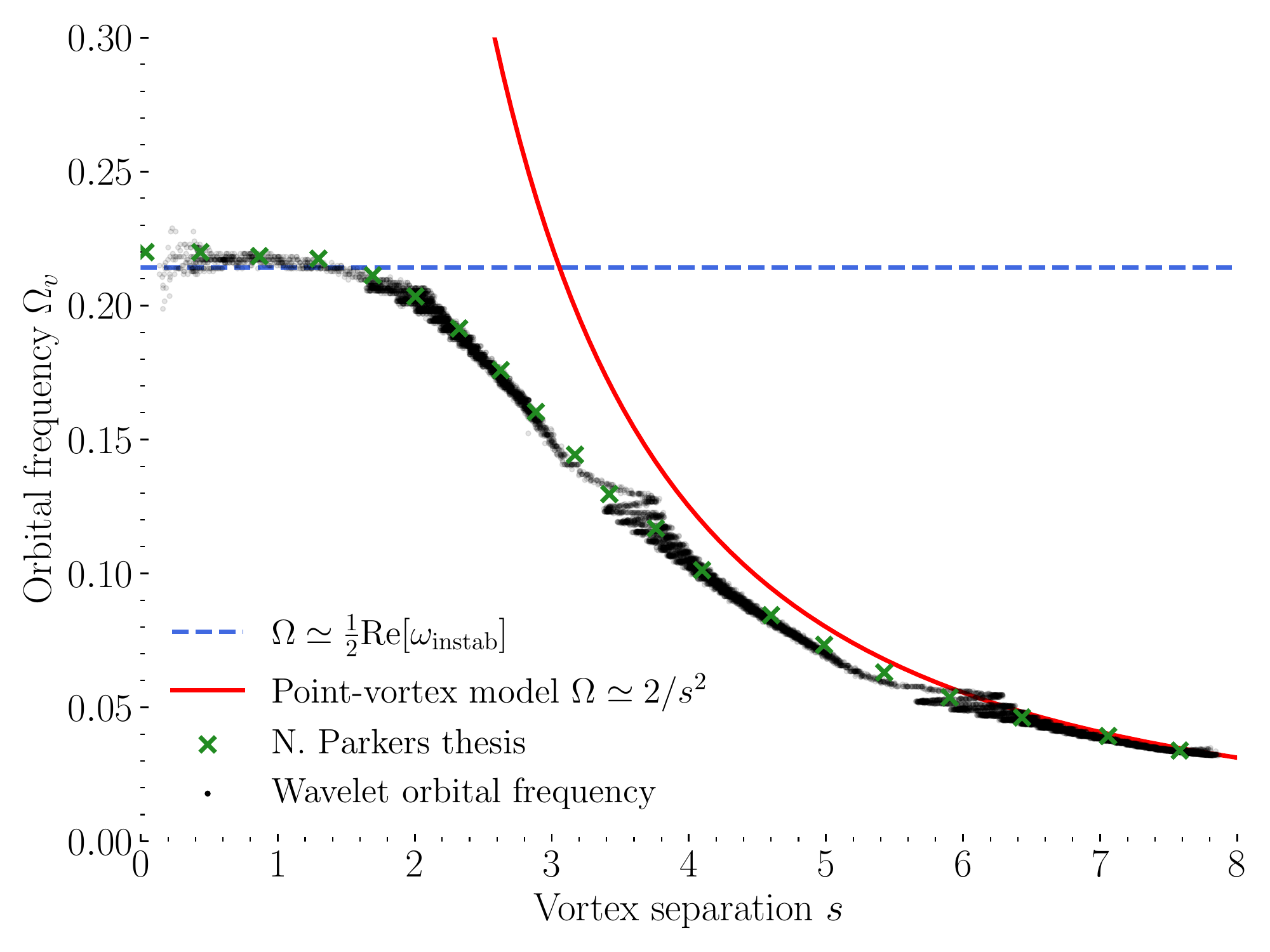}
\caption{The orbital frequency $\Omega$ of the two singularities as a function of distance $s$ between them (black line). At small $s$, the singularities are proto-vortices with $\Omega$ corresponding to frequency of the unstable cavity mode (blue dashed line). At large $s$, the singularities are true vortices which obeys the expected Hamiltonian point vortex dynamics (red line).
We compare our data to a function used by Parker \cite{parker2004numerical} (green crosses) who also considered the dynamics of nearby vortices.}
\label{fig:damped}
\end{figure}

\noindent Dissipation is introduced in the GPE by using the phenomenological damping parameter $\gamma$ \cite{Cockburn_2009,Proukakis_2008},
\begin{equation} \label{GPEgamma}
i\partial_t\Psi = \left(1-i\gamma\right)\left(-\tfrac{1}{2}\nabla^2 + V(\mathbf{x})-1 +|\Psi|^2\right)\Psi.
\end{equation}
The presence of dissipation damps out the oscillations of the two proto-vortices, which spiral away from each other and develop separate core regions. At this point we recover the well-known configuration of two point-vortices of the same sign, which rotate around each other with an orbital frequency inversely proportional to the square of the vortex separation $s$,
as shown in Fig.~\ref{fig:damped}.
The fact that the point-vortex description predicts a divergence as $s\to 0$, signalling the break-down of the model, is a consequence of the overlapping vortex cores.
In this regime, the system is better described as a perturbation (cavity mode) on an $\ell=2$ vortex background.
This mode enters the linear regime in the limit $s\to 0$ and we see that the orbital frequency of the singularities tends to half of the oscillation frequency of the instability. The half is because two nearby singularities make a dipole perturbation, i.e. $m=2$, hence it takes twice the time for a given peak of the $m=2$ mode to return to it's original position.

Another interesting feature of Fig.~\ref{fig:damped} is the sudden oscillations in $s$ as the orbital frequency decreases. This occurs when the cavity mode reaches the correct frequency to couple to another phonon in the system. There are two such oscillations in Fig.~\ref{fig:damped} because at $r_B=25$, there are two phonons with lower frequency than the cavity mode. Hence for a small system where the cavity mode initially couples to the lowest phonon, these sudden additional oscillations would not occur.

\begin{figure*} 
\centering
\includegraphics[width=\linewidth]{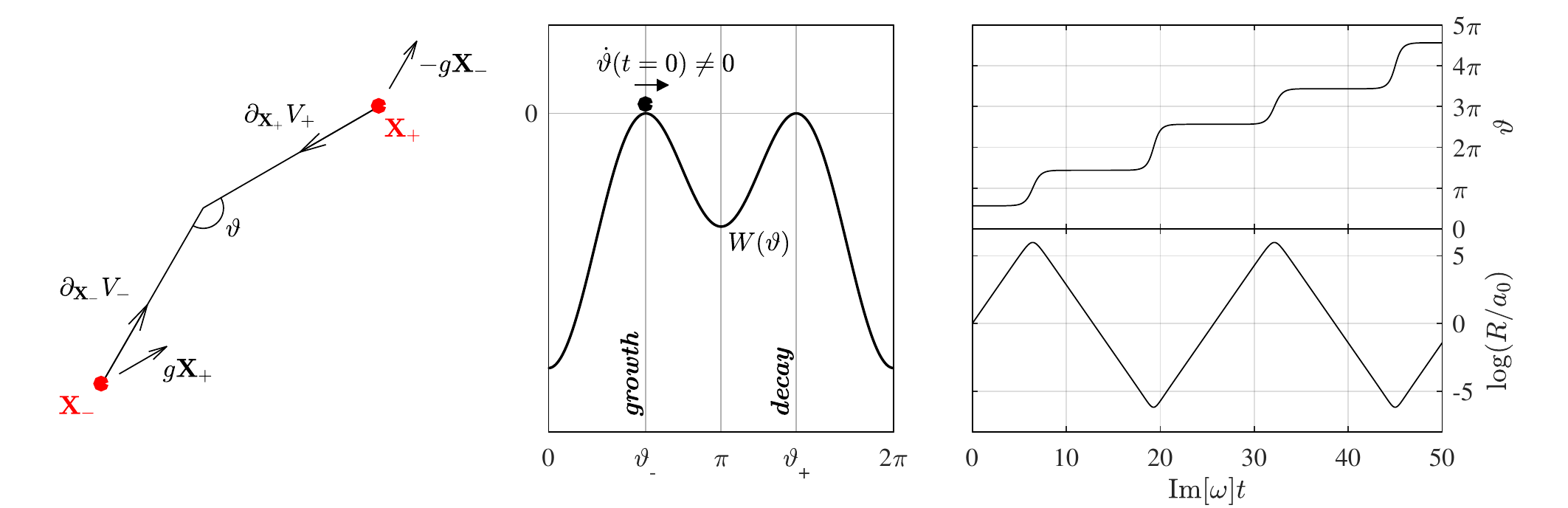}
\caption{\textbf{Left panel:} A schematic of the two interacting oscillators described by \eqref{model1}, where $\mathbf{X}_+$ represents a positive energy phonon and $\mathbf{X}_-$ represents the negative energy oscillation of the vortex (cavity mode). \textbf{Central panel:} In the limit that the central oscillator frequency $\Omega$ is much larger than other scales in the problem, the dynamics can be re-expressed as a non-linear oscillator $\vartheta$ moving through periodic potential $W$ with two wells. \textbf{Right panel:} When $\vartheta$ evolves as it rolls along $W$, the original amplitude $R=|\mathbf{X}_\pm|$ switches between exponential growth and decay, mimicking the observed behaviour in our simulations.}
\label{fig:supp}
\end{figure*}

\section{3.~Two-oscillator model}
The switch from exponential growth to decay observed in our simulations is characteristic of two oscillators (with opposite sign energies) interacting under the influence of a non-linearity.
We illustrate this using a simplified model with the following Lagrangian,
\begin{equation} \label{model1}
\begin{split}
L = & \ \tfrac{1}{2}\dot{\mathbf{X}}_+^2-V_+ - \tfrac{1}{2}\dot{\mathbf{X}}_-^2+V_- - g\mathbf{X}_+\cdot\mathbf{X}_-, \\
V_\pm = & \ \tfrac{1}{2}(\Omega^2\pm\sigma)\mathbf{X}_\pm^2 - \tfrac{1}{4}c_\pm\mathbf{X}_\pm^4,
\end{split}
\end{equation}
which describes two particles at $\mathbf{X}_\pm = (X_\pm,Y_\pm)$ oscillating about the coordinate origin with energies $H_\pm = \pm(\tfrac{1}{2}\dot{\mathbf{X}}_\pm^2+V_\pm)$ respectively.
The interaction energy between the two particles is $g\mathbf{X}_+\cdot\mathbf{X}_-$.
We set $c_+=0$ and $c_-=\varepsilon>0$ so that the non-linearity only affects the particle located at $\mathbf{X}_-$.

To see that this model captures the essential features of the vortex evolution, we consider two limiting cases.
When $g=0$, the two particles oscillate at fixed radii as $\mathbf{X}_\pm\sim e^{-i\omega_\pm t}$ with frequencies $\omega_+=\{\Omega^2+\sigma\}^\frac{1}{2}$ and $\omega_-=\{\Omega^2-\sigma-\varepsilon\mathbf{X}_-^2\}^\frac{1}{2}$. 
Here $\mathbf{X}_-$ mimics the way the orbital frequency of two point vortices decreases as the distance between them grows, whilst $\mathbf{X}_+$ mimics the phonon whose frequency (determined predominantly the system size) stays nearly constant.
Next, when $\varepsilon=0$, both particles oscillate about the origin as a linear superposition of the frequencies $\omega=\{\Omega^2\pm\sqrt{\sigma^2-g^2}\}^\frac{1}{2}$.
Notice that when the oscillator spacing is smaller than the coupling, $|\sigma|<|g|$, this will have unstable solutions, mimicking the behaviour in Fig.~\ref{fig:1} where instabilities occur if a $\mathcal{N}>0$ mode comes close to a $\mathcal{N}<0$ mode in the $\omega$-plane.

The full model with $\varepsilon,g$ non-zero has an elegant solution in the regime $\Omega\gg\sigma,g,\varepsilon\mathbf{X}_-^2$ (which is the relevant one for the vortex where $\mathrm{Re}[\omega]$ is an order of magntiude larger than $\mathrm{Im}[\omega]$).
Defining the complex variable $Z_\pm = e^{i\Omega t}(X_\pm+iY_\pm)$ and rescaling such that $\Omega=1$ and $\varepsilon=2$, the Lagrangian becomes,
\begin{equation} \label{model_lagr}
\begin{split}
L = & \ \mathrm{Im}[Z_+\dot{Z}^*_+ - Z_-\dot{Z}^*_-] - \tfrac{1}{2}\sigma(|Z_+|^2+|Z_-|^2) \\
& \qquad - g\mathrm{Re}[Z_+Z_-^*] - \tfrac{1}{2}|Z_-|^4. 
\end{split}
\end{equation}
This has a conserved charge $\mathcal{Q}=|Z_+|^2-|Z_-|^2$ analogous to $\mathcal{N}$ for the vortex.
For exponentially growing/decaying modes, this is only conserved when $\mathcal{Q}=0$, implying $Z_\pm = |Z_\pm|e^{-i\varphi_\pm}$ with $|Z_\pm|=R$.
In terms of $R$ and the phase difference $\vartheta=\varphi_+ - \varphi_-$, we get,
\begin{equation}
L = R^2(\dot{\vartheta} - \sigma -g\cos\vartheta - \tfrac{1}{2}R^2),
\end{equation}
which leads to the following equations of motion,
\begin{equation} \label{model_eqns}
    \dot{\vartheta} = \sigma+g\cos\vartheta+R^2, \qquad \dot{R} = \tfrac{1}{2}g R\sin\vartheta.
\end{equation}
These can then be combined into a single equation for $\vartheta$,
\begin{equation}
\ddot{\vartheta} + W'(\vartheta) = 0, \qquad W(\vartheta) = -\tfrac{1}{2}(\sigma+g\cos\vartheta)^2,
\end{equation}
which describes a particle on $\mathbb{S}^1$ moving through a potential $W$.
When the linear equations are unstable, $W$ has two maxima $\vartheta_\pm = \pi\pm\cos^{-1}(\sigma/g)$.
By considering the phase of the solutions when $\varepsilon=0$, one finds that the stationary solutions $\vartheta=\vartheta_-$ and $\vartheta=\vartheta_+$ correspond respectively to the linearly growing and decaying modes.
However, in the non-linear case, \eqref{model_eqns} tells us that a non-zero initial amplitude in the instability means $\dot{\vartheta}\neq 0$, making $\vartheta$ roll all the way over the two peaks. 
Once $\vartheta(t)$ is known, the amplitude can be found by solving the equation in \eqref{model_eqns} for $\dot{R}$,
\begin{equation}
    R(t)\sim\exp\left(\frac{1}{2}\int g\sin\vartheta(t)dt\right),
\end{equation}
which tells us that $R$ switches between exponentially growing and decaying solutions as $\vartheta$ changes by $2\pi$.
This behaviour is illustrated in Fig.~\ref{fig:supp}.

The key feature of this model that renders it so simple is that is contains only two interacting modes.
This is what leads to the periodicity of the growth-decay cycles in Fig.~\ref{fig:supp}.
When other modes are taken into account (i.e. the other $\omega$ and $m$ modes around the vortex) energy can be dissipated into these channels and we would expect these cycles to lose perfect periodicity.
This imperfect periodicity appears to be what we see in Fig.~\ref{fig:2} of the main text.


\end{document}